\documentclass[aps,twocolumn,pre,showpacs,eqsecnum]{revtex4}
\usepackage{graphpap}
\usepackage[dvips]{graphicx}
\usepackage[dvips]{graphics}
\usepackage{color}

\begin{document}

\title{Disorder induced Coulomb gaps in graphene constrictions with different aspect ratios}
 \author{B. Terr\'es$^{1,2}$, J. Dauber$^{1}$, C. Volk$^{1,2}$, S. Trellenkamp$^2$, U. Wichmann$^1$, and C. Stampfer$^{1,2}$}
 \affiliation{
$^1$JARA-FIT and II. Institute of Physics B, RWTH Aachen University, 52074 Aachen, Germany, EU\\
$^2$Institute of Bio and Nanosystems, Forschungszentrum J\"ulich, 52425 J\"ulich, Germany, EU
}

\date{ \today}

 \begin{abstract}
We present electron transport measurements on lithographically defined and etched
graphene nano-constrictions with different aspect ratios, including different
length ($l$) and width ($w$). A roughly length-independent disorder induced effective energy
gap can be observed around the charge neutrality point.
This energy gap scales inversely with the width even in regimes where the
 length of the constriction is smaller than its width ($l<w$). In very short
constrictions we observe both resonances due to localized states or charged islands
and an elevated overall conductance level (0.1 - 1$e^2/h$), which is strongly length-dependent in the gap region. This makes
very short graphene constrictions interesting for highly transparent graphene tunneling
barriers.

 \end{abstract}

 \maketitle

\newpage
Graphene nanoribbons and constrictions~\cite{che07,han07,dai08,wan08,mol09,sta09,tod09,liu09,mol10,gal10,han10} are promising candidates
to overcome the gapless nature of graphene~\cite{gei07}.
Graphene, a truly two-dimensional (2D) semi-metal with unique electronic properties, is becoming increasingly interesting for high mobility ultra-fast nanoelectronics~\cite{kat07}.
 However, the missing
band gap in graphene makes it difficult to realize graphene switches, transistors,
and logic devices in general.
By tailoring graphene into narrow ribbons or constrictions, the formation of
an energy gap has been shown.

 Graphene constrictions have been successfully demonstrated to act as tunneling
barriers in single electron transistors~\cite{sta08b}, quantum dots~\cite{pon08,gue09} and more recently in
double quantum dots~\cite{mol09a,mor09,liu10}. The nature of the observed energy gaps has been discussed
in great detail. From the theoretical point of view, different models have been put forward
to describe the observed energy gap including quasi-1D Anderson localization~\cite{eva08,que08,gun08,lhe08,muc09,mar09},
percolation~\cite{ada08} or Coulomb blockade on a series of localized states~\cite{sol07}.
Experimentally, the nanoribbon width-dependence of the energy gap has been studied by
a number of groups and it has been found that the gap size is mainly given by the
charging energy of small isolated islands forming in the nanoribbon due to
disorder~\cite{sta09,tod09}.
However, only recently has the influence of the nanoribbon length, $l$, been
investigated systematically for nanoribbons with width $w<40$~nm~\cite{gal10,han10}. So far, the
experimentally studied graphene nanostructures exhibited an aspect ratio of $l>w$,
such that
the width was
the smallest and dominating length scale. 

Here, we present a systematic study of electronic transport in graphene constrictions with different aspect
ratios, including both different width and length. In particular, we focus on graphene
constrictions in the regime of $l \approx w$ and $l<w$ [see Fig.~1(a)].
We show that in these regimes,
the Coulomb charging energy $E_g$ of the smallest disorder-induced charged island in the constriction~\cite{sta09,tod09}
is still predominantly determined by the width.
%
However, the overall conductance level is strongly influenced by the length of the constriction.
Most interestingly, we show that in very short constrictions, Coulomb blockade diamonds with high
transparency can be observed. 

The samples have been fabricated from graphene that has been mechanically exfoliated from natural bulk graphite and
deposited onto 295~nm SiO$_2$ on a highly p-doped Si substrate.
Raman spectroscopy is used to verify the single-layer character of the investigated graphene nanostructures~\cite{dav07a}.
Electron-beam (e-beam) lithography is
used to pattern the etch mask ($\approx$~110 nm of e-beam resist) for structuring the graphene devices.

 \begin{figure}[hbt]\centering
\includegraphics[draft=false,keepaspectratio=true,clip,%
                   width=0.99\linewidth]%
                   {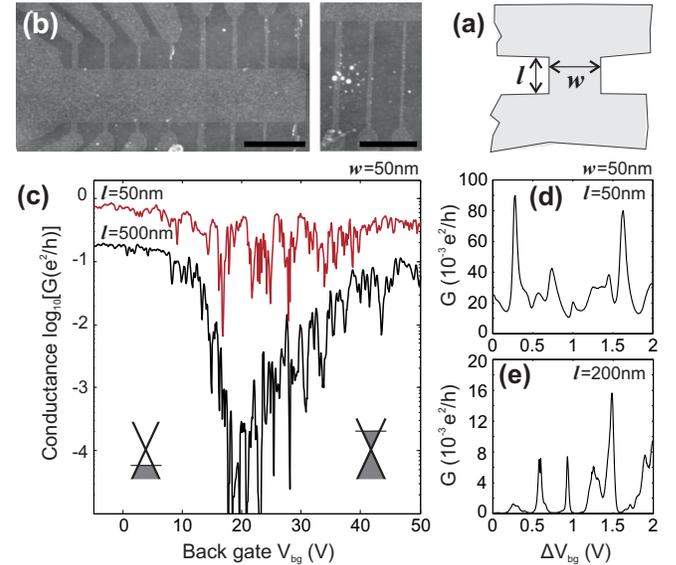}
\caption[FIG1]{(color online)
(a) Illustration of a constriction with length $l$ and width $w$.
(b) Scanning force microscope images of etched graphene constrictions with different aspect ratios. The length of the scale bars is 1~$\mu$m.
(c) Back-gate characteristics at $V_b = 500~\mu$V for two graphene constrictions with width
$w=50$~nm and different length $l=50$~nm (red curve) and $l=500$~nm (black curve).
(d,e) Conductance peaks in the transport gap region for two different constrictions, $w=l=50$~nm (d) and $w=l/4=50$~nm (e) at $V_b = 100~\mu$V.}
\label{trdansport}
\end{figure}

Reactive ion etching based on an Ar/O$_2$ plasma is employed to remove unprotected graphene.
Scanning force microscope images of etched graphene nanostructures after removing the residual resist are shown in Fig.~1(b).
Finally, the graphene devices are contacted by e-beam patterned 5~nm Cr and 50~nm Au electrodes.

\begin{figure}[t]\centering
\includegraphics[draft=false,keepaspectratio=true,clip,%
                   width=0.99\linewidth]%
                   {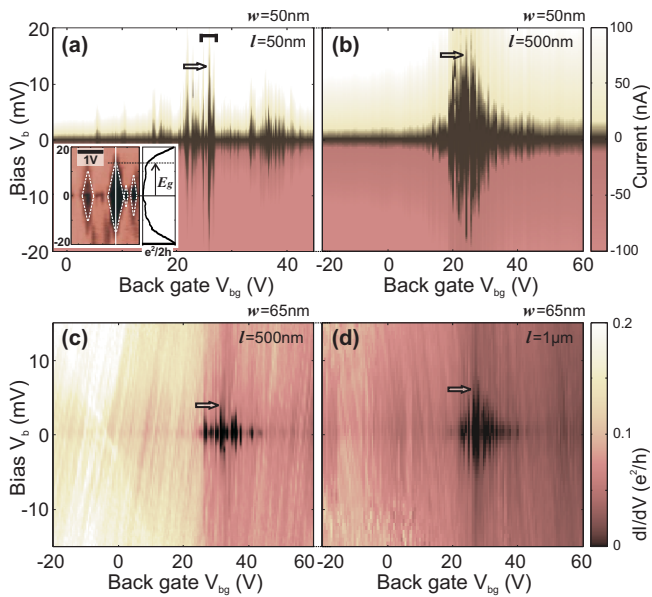}
\caption[FIG2]{
(color online) (a,b) Color plot of the source-drain current as a function
of $V_{bg}$ and $V_{b}$ for two constrictions with
$w=50$~nm and $l=50$~nm (a) and $l=500$~nm (b). The inset shows
differential conductance measurements of the region where the transport
gap is most pronounced [see black bar in panel (a)]. A cross-section through the center
of the largest diamond is shown on the right side of the inset, highlighting the definition
of the gap $E_g$. (c,d) Differential
conductance maps for graphene constrictions with
$w=65$~nm and length $l=500$~nm (c) and $l=1~\mu$m (d). 
}
\label{expresults}
\end{figure}

The measurements have been performed in a pumped $^4$He system at $T\approx$~1.3~K. We have measured the current through the two-terminal constrictions by applying a small symmetric DC bias voltage $V_{b}$. The differential conductance has been measured directly by low-frequency lock-in techniques by adding an AC bias of 100~$\mu$V.

In Fig.~1(c) we show the conductance $G$ as function of back gate (BG) voltage $V_{bg}$ for two 50~nm wide
constrictions with different length $l=50$~nm and $l=500$~nm ($V_b$ = 500~$\mu$V). The transport gap,
 i.e. the $V_{bg}$ region of suppressed conductance (roughly between 17 and 30~V), separates hole from electron transport
 as indicated by the two lower inserts in Fig.~1(c)~\cite{com01}.
Whereas for the $500$~nm constriction the conductance is strongly suppressed in this
regime (down to $10^{-5} e^2/h$), we observe a significantly
increased conductance for the shorter constriction. Nevertheless, a number of pronounced resonances reaching lower conductance
values mark the gap region in rather good agreement with the measurement on the longer
constriction [compare also Fig.~2(a) and 2(b)].
A close-up of such resonances is shown in Fig.~1(d). These reproducible
resonances have been taken at $V_b=100~\mu$V within the BG range of minimum conductance.
We observe neither complete pinch-off nor strong Coulomb blockade behavior, as
is found for longer constrictions with the same width. For example, in Fig.~1(e) we show
data recorded on a $200$~nm long ($w=50$~nm) constriction.
The overall conductance level differs significantly, however the typical $V_{bg}$ spacing between the
resonances is comparable.

\begin{figure}[t]
\centering
\includegraphics[draft=false,keepaspectratio=true,clip,%
                   width=0.90\linewidth]%
                   {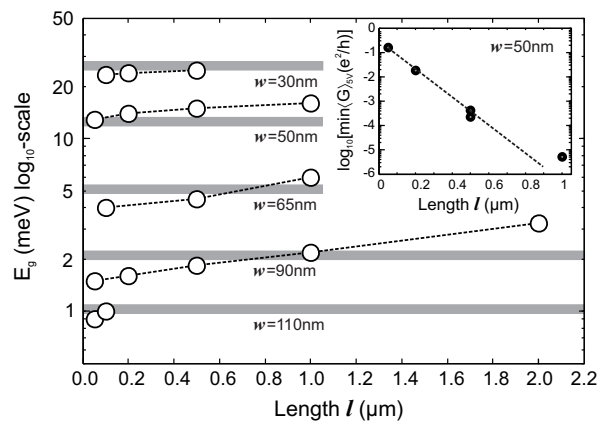}
\caption[FIG3]{
Energy gap, $E_g$, as a function of length $l$ for a number of
different graphene constrictions with different width $w$ (see labels
and data points connected by dashed lines). The horizontal gray lines
are given by $E_g=a/w e^{-bw}$, where $a=2$~eV and $b=0.026$~nm$^{-1}$ taken from ref.~\cite{mol10}. The inset
shows the minimum value of the running averaged conductance $G$ over $5$~V on back gate voltage.
}
\label{s2}
\end{figure}

In Figs.~2(a),~2(b) we show 2D-plots of current as function of bias and back gate voltage for the two
$50$~nm wide constrictions with $l=50$~nm (a) and $500$~nm (b).
 In good agreement with earlier
studies~\cite{han07,mol09,sta09,tod09,liu09,mol10,gal10,han10} we find
regions of suppressed current with an extension of $E_g/e$ in the $V_b$ direction (see arrows).
By plotting higher-resolution measurements of the differential conductance, one can
see that the region of suppressed current is composed of individual diamonds [see inset in Fig. 2(a)].
The energy gap $E_g$ corresponds to the charging energy of the largest diamond~\cite{sta09,tod09}, as highlighted in the inset
[see arrow in Fig.~2(a)]. Interestingly, this energy scale only
weakly depends on the constrictions length.
By comparing Figs.~2(a) and~2(b) we observe only a small difference in $E_g$ between
the $50$~nm ($E_g$ = 13~meV) and $500$~nm ($E_g$ = 15~meV) long constriction, allowing the conclusion that the smallest island or
localized state
is predominantly a function of the width, $w$.
%
%
%
The main difference can be found in the back gate coverage of the
observed gaps. The shorter the constriction, the fewer islands [or localized (edge) states] are in
the constriction, leading to fewer charging events.
Consequently, the current is suppressed in smaller and fewer
gate voltage regimes. However, the smallest island is found to be roughly length-independent (this is
also true for $l<w$). The definition of the transport gap in BG voltage~\cite{mol09,mol10}, used as
figure of merit in earlier work, is hard to define for very short constrictions since it is considered
to strongly depend on the disorder potential~\cite{sta09,tod09,gal10}, which becomes very
sample dependent, due to lack of averaging.

Similar behavior can also be seen for a $65$~nm wide, $500$~nm and $1~\mu$m long constriction,
as shown in Figs.~2(c) and~2(d), respectively.
Here we plot the differential conductance as a function of $V_b$ and $V_{bg}$.
In both measurements, a maximum charging energy  of around 4.5 and 6~meV can be observed (see arrows).
\begin{figure}[t]
\centering
\includegraphics[draft=false,keepaspectratio=true,clip,%
                   width=0.99\linewidth]%
                  {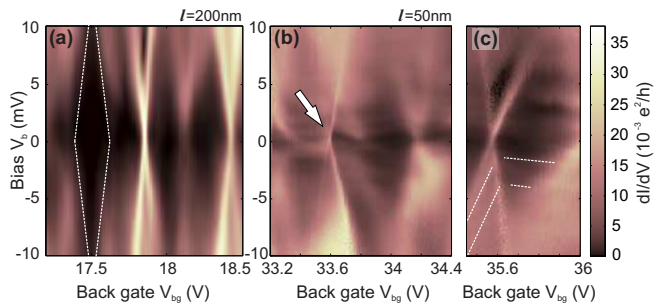}
\caption[FIG4]{
(color online)
Differential conductance versus BG voltage and bias voltage of (a) 200 nm and (b,c) 50 nm long
graphene constriction with a width of 50 nm. (a) Distinct diamonds of fully suppressed conductance can be observed
(see white dashed line). (b,c) Similar data taken on a 50 nm long constriction. Diamonds at an elevated conductance level can be observed. Sharp resonances [white arrow in panel (b)] and features of increased conductance inside the diamonds [horizontal dashed lines in panel (c)] can be observed.}
\label{s3}
\end{figure}
In total, we studied roughly 20 graphene constrictions on 3 different samples. In Fig.~3 we summarize the extracted
energy gaps $E_g$ as a function of the length for 5 different widths~\cite{com02}.
It can be seen that $E_g$ strongly depends on the width and we find good agreement with earlier
experiments and theoretical models~\cite{sol07,han10}. In particular, we compare our results with the model
from Sols et al.,~\cite{sol07} where the energy gap is approximated by $E_g=a/w e^{-bw}$, (see gray bars in Fig. 3), see also~\cite{han07,sol07,mol10}.
In contrast to the weak $E_g$-length dependence, we observe a rather strong length dependence of the minimum conductivity, as shown by the inset in Fig.~3. The minimum value of the running averaged conductance decreases exponentially with increasing constriction length.
This fits well with the scenario where localizations or tunneling processes are dominating the transport
through the constrictions.
Moreover, it shows that by making graphene constrictions very short, we can obtain conductance level close to $e^2/h$.

More insights on the transparency of the shorter constrictions is gained by focusing on a small BG voltage range as shown in Fig.~4.
Here, we show high-resolution differential conductance $dI/dV_b$ plots for different constriction lengths with constant width ($w$=50~nm).
Measurements on a 200~nm long constriction [Fig.~4(a)] show well-distinguishable diamonds of suppressed conductance in good agreement with earlier studies. In shorter constrictions [$w$=50~nm; Figs.~4(b),4(c)] we observe
diamonds where the conductance is not fully suppressed, most likely because of strong coupling to localized states or isolated charged islands.
 Moreover, we observe faint lines of increased conductance inside diamonds aligned with features outside the
diamonds, which might be due to inelastic co-tunneling [see e.g. dashed lines in Fig.~4(c)].
According to recent experiments
by Han et al.~\cite{han10} the average hopping length $L_c$ is in the range of $w \lesssim L_c < 2w$. Consequently,
it is likely that for very short constrictions $L_c$ exceeds $l$ ($l < w \lesssim L_c$), so that transport
in short constrictions is no longer effectively 1D, resulting in several transport channels in the constrictions. This would account for
(i) the strong increase in conductance and
(ii) the appearance of very sharp features in the conductance [see e.g. arrow in Fig.~4(b)],
which could result from interference effects or Fano resonances.


In summary, we presented transport measurements on
graphene nano-constrictions with different aspect ratios.
We showed that the strongly width-dependent disorder induced energy gap is roughly length-independent, whereas
the overall conductance level depends strongly on the length. In the gap
region of very short graphene constrictions, a conductance level close to 0.1e$^2$/h can be reached,
making these structures
potential candidates for exploring Fano resonances and Kondo physics in graphene nanostructures.

{Acknowledgment ---}
The authors wish to thank C. Barengo, P. Studerus, M. Morgenstern, D. Gr\"utzmacher, P. Kordt, J. Schirra and S. Gustavsson
 for their support and help in setting up our new lab.
We thank A. Steffen, R. Lehmann and J. Mohr for the help on the sample fabrication.
Discussions with H.~L\"uth, T. Heinzel, T. Ihn, J. G\"uttinger
and support by the JARA Seed Fund and SPP-1459 are gratefully
acknowledged.


\begin{thebibliography}{99}
\bibitem{com01}
All samples are slightly p-doped most likely due to residual resist on the graphene structures.
\bibitem{com02}
Since the extracted gap $E_g$ is strongly disorder dependent the error bars are around 5-10\%, in agreement with earlier studies. For clarity error bars are not shown explicitly in Fig.~3.
\end{thebibliography}
\end{document}